\documentclass[12pt]{article}
\usepackage{fullpage,epsfig,graphics,amsbsy,amssymb}
\usepackage{graphicx}
\usepackage{amsfonts}
\usepackage{amssymb,amsmath,amscd}

\newtheorem{proposition}{Proposition}

\newtheorem{example}[proposition]{Example}

\newcommand{\bea}{\begin{eqnarray}}
\newcommand{\eea}{\end{eqnarray}}
\newcommand{\beq}{\begin{eqnarray}}
\newcommand{\eeq}{\end{eqnarray}}
\newcommand{\nn}{\nonumber}

\def\M{{\cal M}}
\def\N{{\cal N}}

\begin{document}

\thispagestyle{empty}
\begin{flushright} \small
UUITP-22/09  \\
\end{flushright}
\smallskip
\begin{center} \LARGE
{\bf  2D and 3D topological field theories\\
for generalized complex geometry }
\\[12mm] \normalsize
{\bf Alberto~S.~Cattaneo$^a$,  Jian~Qiu$^b$ and Maxim Zabzine$^b$} \\[8mm]
{\small\it
$^a$Institut f\"ur Mathematik, Universit\"at Z\"urich-Irchel\\
 Winterthurerstrasse 190, CH-8057 Z\"urich, Switzerland\\
 ~\\
$^b$Department of Physics and Astronomy,
    Uppsala university,\\
    Box 516,
    SE-75120 Uppsala,
    Sweden\\}
\end{center}
\vspace{10mm}
\centerline{\bfseries Abstract} \bigskip
Using the AKSZ prescription  we construct 2D and 3D topological field theories
 associated to  generalized complex manifolds.  These models can be thought of
  as 2D and 3D generalizations of A- and B-models. Within the BV framework we show
   that the 3D model on a two-manifold cross an interval 
can be reduced to the 2D model.

\noindent

\eject
\normalsize


\eject

\section{Introduction}

Recently, generalized complex geometry  has attracted considerable interest both in the physics
and mathematics communities.   Generalized complex geometry has been introduced by Hitchin \cite{Hitchin:2004ut}
and further developed by Gualtieri \cite{Gualtieri:2003dx}  as a notion which unifies symplectic and complex geometries.
 At the same time   generalized complex geometry can be thought of as a complex analogue of the Dirac
geometry introduced by Courant and Weinstein in \cite{courant, courant-weinstein}.

In this work we discuss the Batalin--Vilkovisky (BV) formulation \cite{Batalin:1981jr} of two- and three-dimensional topological
sigma models with target a generalized complex manifold.  Generalized complex geometry
 has a simple description \cite{grabowski} in the language of graded manifolds. This will enable us to  use the
Alexandrov--Kontsevich--Schwarz--Zaboronsky  (AKSZ) prescription \cite{Alexandrov:1995kv} for the construction
of solutions to the classical master equation.
 We study the relation between 3D and 2D models within the  BV framework. Naturally
 our results have a wider interpretation in the context of general 3D and 2D AKSZ models.
  This work  contains only the construction of the models; issues such as gauge fixing, localization
   and the calculation of correlators are left for another more technical paper \cite{CMQZ}.

 Let us comment on the literature and on the relations between our and others' work.
Different 2D and 3D versions of 
  topological sigma models for generalized complex structures
were discussed previously within the  BV formalism. To mention some, there are the
  two dimensional Zucchini model  \cite{Zucchini:2004ta}, the  three dimensional Ikeda models \cite{Ikeda:2004cm, Ikeda:2006pd}
   and the Pestun model \cite{Pestun:2006rj}.  These models are interesting on their own.
         Our main intention here is to show that the powerful AKSZ framework
    produces the simple and unique 2D and 3D models associated to generalized complex geometry.
     Moreover, 2D and 3D models are related to each other in a rather canonical way.

The article is organized as follows: Section \ref{AKSZ} contains a brief review of the AKSZ construction
of  solutions to the classical master equation. In  Section \ref{AKSZ1and2} we review the AKSZ models
 with target a symplectic graded  manifold of degree $1$ or $2$.   Section \ref{AKSZ_for_GCG}
 recalls the description of generalized complex geometry in terms of graded manifolds. This enables us to
  construct two- and three-dimensional AKSZ models.  In  Section \ref{reduction2D}
   we discuss the relation between these models. The main idea is to use  Losev's trick \cite{losev},
    the partial integration of a subsector of the theory.  Section \ref{summary} gives a summary
     and provides an outlook to forthcoming work. At the end of the paper we present two technical
      Appendices with the explicit formulas describing generalized complex geometry in the language
       of graded manifolds.

\section{The AKSZ-BV formalism}
\label{AKSZ}

The BV formalism \cite{Batalin:1981jr} is a
powerful tool in the quantization of an action functional that is
degenerate (e.g. due to gauge equivalence). This procedure embeds
the space of fields into the so called BV manifold,
which is equipped with an odd symplectic structure and thereby an
odd BV bracket $\{\cdot,\cdot\}$. The original action is enlarged
to a new action $S$ satisfying the so called master equation
$\{S,S\}=0$. One then chooses a Lagrangian submanifold inside the
BV manifold; the original path integral is now replaced by the
integration of $S$ over this Lagrangian submanifold. The
geometrical essence of this procedure was expounded by Schwarz
\cite{Schwarz:1992nx}, which reformulated the BV formalism as the
'$PQ$-structure' on a supermanifold. The $P$-structure is just the
symplectic structure and the $Q$-structure is a nilpotent vector
field $Q$ that corresponds to $\{S,\cdot\}$ in the BV case.

In this Section we review the AKSZ construction \cite{Alexandrov:1995kv} of  solutions of the classical master equation
within BV formalism.  We closely follow the presentation given in \cite{Roytenberg:2006qz} and use the language
 of graded manifolds which are sheaves of $\mathbb{Z}$-graded commutative algebras over a smooth manifold;
  for further details the reader may consult \cite{Voronov:2001qf}. We consider both the real and complex cases and treat
   them formally on equal footing. However, in the complex case additional care is required
    (see  \cite{Alexandrov:1995kv} for further details).

The AKSZ solution of the classical master equation is defined starting from the following data:

\medskip
\noindent{\bf The source}: A graded manifold $\N$ endowed with a homological vector field $D$ and
a measure $\int\limits_\N\mu$ of degree $-n-1$ for some positive integer $n$ such that the measure is
invariant under $D$.

\noindent{\bf The target}:  A graded symplectic manifold $(\M,\omega)$ with $\deg(\omega)=n$ and a
homological vector field $Q$ preserving $\omega$.  We require that $Q$ is Hamiltonian,
i.e. there exists a function $\Theta$ of degree $n+1$  such that $Q=\{\Theta,-\}$.  Therefore $\Theta$ satisfies
the following Maurer--Cartan equation
$$\{ \Theta, \Theta \}=0~.$$

Introduce the (infinite dimensional) graded manifold ${\rm Maps}(\N,\M)$ of maps from ${\cal N}$ to ${\cal M}$.
Its body is the manifold of morphisms from ${\cal N}$ to ${\cal M}$ (i.e., sheaf morphisms of the sheaves
describing the two graded manifolds). A soul is added to allow for morphisms parametrized by other graded
manifolds: namely,  ${\rm Maps}(\N,\M)$ is uniquely characterized by the property that morphisms
from $\mathcal{P}\times\N$ to $\M$ are the same as morphisms from $\mathcal{P}$ to ${\rm Maps}(\N,\M)$
for any graded manifold $\mathcal P$. By abuse of language we will often speak of maps from ${\cal N}$ to ${\cal M}$
and write $\N\longrightarrow\M$
when referring to constructions involving ${\rm Maps}(\N,\M)$. 

With our choices for $\N$ and $\M$, ${\rm Maps}(\N,\M)$ is naturally
  equipped with an odd symplectic structure; moreover, $D$ and $Q$ can be interpreted as homological vector
   fields on ${\rm Maps}(\N,\M)$ that preserve this odd symplectic structure. The AKSZ solution $S_{BV}$
     is the Hamiltonian for the homological vector field $D+Q$  on ${\rm Maps}(\N,\M)$ and thus it satisfies 
      the classical master equation automatically.

Let us provide some details for this elegant construction.
We denote by $\Sigma$ and $M$ the underlying smooth manifolds to $\N$ and $\M$ respectively.
We choose a set of coordinates $X^A=\{x^\mu;\psi^m\}$ on
the target $\M$, where $\{x^\mu\}$ are the coordinates for an open $U\subset M$ and $\{\psi^m\}$
are the coordinates in the formal directions.  We also choose  coordinates $\{\xi^\alpha; \theta^a\}$ on the source
${\cal N}$, where   $\{\xi^\alpha\}$ are the local coordinates on $\Sigma$ and $\{ \theta^a\}$ are the coordinates in the formal
 directions of $\N$. 
We then collect local  coordinates on ${\rm Maps}(\N,\M)$ into the  superfield $\Phi$,
\begin{equation}\label{superfield}
\Phi^A= \Phi_0^A(u) +\theta^a  \Phi_{a}^A(u) + \frac{1}{2}\theta^{a_2}   \theta^{a_1}\Phi_{a_1a_2}^A(u)  + \ldots,
\end{equation}
where $\Phi^A_0$, $\Phi^A_a$, $\Phi^A_{a_1a_2}$, \dots (the coordinates on ${\rm Maps}(\N,\M)$), 
are functions on  $\Phi_0^{-1}(U)$. They are assigned a degree such that $\Phi^A$ has degree equal to the
degree of $X^A$. 

The symplectic form $\omega$ of degree
$n$ on $\M$ can be written in Darboux coordinates as $\omega = dX^A \omega_{AB} dX^B$.
Using this form we define the symplectic form of degree $-1$ on ${\rm Maps}(\N,\M)$  as
\begin{equation}
\label{P_structure}
\omega_{BV} = \frac{1}{2} \int\limits_\N \mu ~~\delta \Phi^A ~\omega_{AB}~ \delta \Phi^B~.
\end{equation}
Thus the space of maps ${\rm Maps}(\N,\M)$ is naturally equipped with the odd Poisson bracket $\{~,~\}$.
Since the space ${\rm Maps}(\N,\M)$ is infinite dimensional we cannot define the BV Laplacian properly. We can only talk
  about the naive odd Laplacian.
However on ${\rm Maps}(\N,\M)$ we can discuss the solutions of the classical master equation. Assuming that
  $\omega$  admits a Liouville form $\Xi$ the AKSZ action then reads
\beq\label{AKSZ-action}
S_{BV}[\Phi]= S_{kin}[\Phi]+ S_{int}[\Phi] = \int\limits_\N \mu ~\left (  \Xi_A(\Phi) D\Phi^A +
(-1)^{n+1} \Phi^*(\Theta)\right )
\eeq
and it solves the classical master equation $\{S_{BV},S_{BV}\}=0$ with respect to the bracket
defined by the symplectic structure (\ref{P_structure}).   Since the measure $\mu$ is invariant under $D$, $S_{kin}$ depends
   only on $\omega$, not a concrete choice of $\Xi$.
 In  particular, using the Darboux coordinates  the first term in (\ref{AKSZ-action}) can be written
   \beq\label{Liouvilleform}
  S_{kin} [\Phi] =  \int\limits_\N \mu ~ \frac{1}{2} \Phi^A\omega_{AB} D\Phi^B~.
 \eeq
 The action (\ref{AKSZ-action}) is invariant under all orientation preserving diffeomorphisms of $\Sigma$
   and thus defines a topological field theory.  The solutions of the classical field equations of (\ref{AKSZ-action}) are graded differentiable maps $(\N, D) \rightarrow (\M, Q)$, i.e. maps which commute with the homological vector fields.

 The standard choice for the source is the odd tangent bundle
$\N=T[1]\Sigma_{n+1}$, for any smooth manifold $\Sigma$ of dimension
$n+1$, with $D=d$ the de~Rham differential over $\Sigma$ and the
canonical coordinate measure $\mu = d^{n+1}\xi ~d^{n+1} \theta \equiv d^{n+1} z$
   \beq\label{w0w000w}
    S_{BV} [\Phi] =\int\limits_{T[1]\Sigma_{n+1}}   d^{n+1}z ~\big(\Xi_A(\Phi) D\Phi^A +(-1)^n \Phi^*(\Theta)\big)~.
   \eeq
    For the rest of the paper we consider only the case when the source is $T[1]\Sigma_{n+1}$.
     However, the more exotic situations are possible, e.g. the holomorphic part of an
      odd tangent bundle etc, see \cite{Qiu:2009zv}.
   Next we consider the case when  $\Sigma_{n+1}$ has
a boundary. For this  we need to impose certain boundary conditions within AKSZ prescription, see
\cite{Cattaneo:2001ys} for details.  In particular
the BV classical master equation for (\ref{w0w000w})  is only satisfied up to total
derivative terms
\bea
&&\{S_{BV},S_{BV}\}=\int\limits_{T[1]\Sigma_{n+1}} d^{n+1}z~
D\big(\Xi_{A}(\Phi)D\Phi^A+(-1)^n \Phi^*(\Theta)\big)\\
&&=\int\limits_{T[1]\partial\Sigma_{n+1}} d^n z~
\big(\Xi_{A}(\Phi)D\Phi^A+(-1)^n\Phi^*(\Theta)\big)~.\nn
\eea
Thus a natural choice for the boundary condition\footnote{Throughout the paper, for the sake of clarity we assume that $\partial \Sigma_{n+1}$ has
a single component. The generalization beyond this case is quite obvious.} is
\beq\label{BC1}
\Phi : T[1]{\partial\Sigma_{n+1}}\rightarrow {\cal L}\subset {\cal M}~,
\eeq
where ${\cal L}$ is a Lagrangian submanifold of the target
${\cal M}$ such that
\beq\label{BC2}
\Xi|_{\cal L} =0~,~~~~~~~~~~~~\Theta|_{\cal L}=0~.
\eeq
 Now with these additional conditions
  the  solution $S_{BV}$
     is the Hamiltonian for homological vector field $D+Q$  on ${\rm Maps}(T[1]\Sigma_{n+1} \rightarrow {\cal M}, T[1]\partial \Sigma_{n+1} \rightarrow {\cal L} )$ and thus it satisfies automatically
      the classical master equation.

Let us make a few concluding remarks.  The advantage of the
  AKSZ construction is that it converts complicated questions into
 a simple geometrical framework. For example, the analysis of the classical observables is straightforward.
   The homological vector field $Q$ on $\M$ defines a complex on $C^\infty(\M)$ whose
cohomology we denote  $H_Q(\M)$. Take $f\in C^\infty(\M)$ and expand $\Phi^*f$
in the formal variables on $\N$
$$\Phi^*f= O^{(0)}(f) +  \theta^a O^{(1)}_a(f) + \frac{1}{2} \theta^{a_2} \theta^{a_1}O^{(2)}_{a_1 a_2}(f)  + \ldots ~.$$
 We denote by $\delta_{BV}$  the
Hamiltonian vector field for $S_{BV}$, which is homological as a
consequence of the classical
  master equation.
The action of $\delta_{BV}$ on $\Phi^*f$ is given by the following expression
$$
\delta_{BV}(\Phi^*f)=\{S_{BV} ,\Phi^*f\} = D\Phi^*f + \Phi^* Qf~.
$$
Thus if $Qf=0$ and $\mu_k$ is a $D$-invariant linear functional on the functions of $\N$ (e.g.,
a representative of an homology class of $\Sigma$), then $\mu_k(O^{(k)}(f))$ is
$\delta_{BV}$-closed and can serve as a classical observable. Therefore $H_Q(\M)$ naturally defines
a set of classical observables in the theory. The classical action (\ref{AKSZ-action}) can be deformed to  first order by
$$ \int\limits_{\N}\mu~ O^{(n+1)}(f)$$
with $f \in H_Q(\M)$.

The gauge fixing in the BV framework corresponds to the choice of a Lagrangian submanifold in the space of fields.
 For a given Lagrangian submanifold we can choose the adapted coordinates with the odd symplectic form
  written as follows
 \beq
  \omega_{BV} = \int\limits_{\cal N} \mu~   \delta \Phi^a ~ \delta\Phi^+_a~,
 \eeq
  such that the Lagrangian is defined by the condition $\Phi^+=0$.
 We expand a master action
formally into a power series  $\Phi^+$
\bea
&&S_{BV} [\Phi ,\Phi^+] =S_{GF}(\Phi)+Q^a (\Phi)  \Phi^+_a + \frac{1}{2}\sigma^{ab} (\Phi) \Phi^+_a \Phi^+_b +\cdots~,\nn\\
&&\{S_{BV},S_{BV}\}=0\Rightarrow Q^a \frac{\partial}{\partial \Phi^a} S_{GF} (\Phi)=0;\
[Q,Q]^a=2\sigma^{ba}\frac{\partial}{\partial
\Phi^b} S_{GF} (\Phi)~. \nn\eea%
Hence  the gauge fixed action $S_{GF}(\Phi)$ has BRST symmetry $Q$
which is nilpotent on shell.  Due to this simple observation it is very easy to analyze
the BRST symmetries of the gauge fixed action.

The AKSZ prescription is algebraic in its nature and thus it  can be generalized  even further, see for
 example \cite{Bonechi:2009kx}.

\section{AKSZ for Symplectic GrMfld of Degree 1 and 2}
\label{AKSZ1and2}

In this Section we review the relevant facts about symplectic graded manifolds (GrMfld) of Degree 1 and 2 with
  nilpotent Hamiltonians of degree 2 and 3 respectively.   The symplectic target
  of degree 1 with  nilpotent Hamiltonian of degree 2 leads to the AKSZ construction of the
   Poisson sigma model \cite{Cattaneo:2001ys} while the symplectic manifold of degree 2
    with nilpotent Hamiltonian of degree 3 leads to the AKSZ construction of the Courant
     sigma model \cite{Roytenberg:2006qz}.

\subsection{Symplectic GrMfld of Degree 1 and 2}

Here we  review the basic facts about symplectic GrMflds of Degree 1 and 2. In particular
we consider some specific examples  which are relevant for our further discussion. Our review is somewhat
 informal and we refer the reader for further details to \cite{roytenberg-thesis, Roytenberg:2002nu}.

When ${\cal M}$ is of degree 1, we denote the coordinates $x,\eta$
with degree 0, 1. The local patches are glued through degree
preserving transition functions. Degree preserving means that the
transition function for the degree 1 coordinate $\eta^A$ must be
linear in $\eta$ and the coefficient of linearity may depend on
the degree zero coordinate $x$ (since we assume that there is no negatively
graded coordinate). One immediately sees that degree 1 GrMflds are
exhausted by $L[1]$, where $L\rightarrow M$ is a vector
bundle. A degree 1 vector field on such a manifold must
have the form
\bea
Q=2\eta^AA_A^{\mu}(x) \frac{\partial}{\partial
x^{\mu}}-f^A_{BC}(x) \eta^B\eta^C\frac{\partial}{\partial\eta^A}\label{Q_d_L}~.
\eea
Requiring $Q^2=0$ puts constraint on the coefficients
\bea&&A_{[A}^{\nu}\partial_{\nu}A_{B]}^{\mu}=A^{\mu}_Cf^C_{AB}~,\nn\\
&&A_A^{\mu}\partial_{\mu}f^D_{BC}+f^D_{AX}f^X_{BC}+\textrm{cyclic in
\small$ABC$}=0~,\label{Lie_Alge_compatible}
\eea
where we use the notation $\partial_\mu = \frac{\partial}{\partial x^\mu}$.
In fact, these data give rise to a Lie algebroid structure \cite{vaintrob}: if one
pick a basis $\ell_A$ for the sections of $L$, then we can
define an anchor map $\pi:L\rightarrow TM:
\pi(\ell_A) \equiv A_A^{\mu}\partial_\mu$, and the structure
function $f^{A}_{BC}$ defines the Lie bracket for the sections of
$L$: $[ \ell_B, \ell_C]=f^A_{BC} \ell_A$ satisfying the
extra condition
$[\ell_A,f \ell_B]=f[ \ell_A, \ell_B]+(\pi(\ell_A)f) \ell_B$.
The second of the equations (\ref{Lie_Alge_compatible}) is the condition for the Jacobi
identity for the Lie bracket, while the first says that the anchor
$\pi$ is a homomorphism between the Lie bracket of $L$ and the Lie
bracket of $TM$. It is also easy to see that $Q$ acts on the
functions $f(x,\eta)$ as the Lie algebroid differential $d_L$ on $\Gamma(\wedge^\bullet L^*)$.
 If one is further restricted to symplectic degree 1 GrMflds, then
one can utilize the symplectic structure to identify the degree 1
coordinate $\eta$ with the fiber coordinate of $T^*M$. In other words degree 1
symplectic manifolds are exhausted by $T^*[1]M$ with symplectic structure
 $$ \omega = d \eta_\mu d x^\mu~,$$
where   $x^\mu$ is coordinate of degree 0 on $M$ and $\eta_\nu$ is the fiber coordinate
 of degree 1. The Hamiltonian of degree 2 is given by the following expression
  \beq\label{Poissonham}
   \Theta = \alpha^{\mu\nu} (x)\eta_\mu \eta_\nu~,
  \eeq
   where $\alpha= \alpha^{\mu\nu} \partial_\mu \wedge \partial_\nu$ is bivector on $M$.
    $\{ \Theta, \Theta\}=0$ if and only if $\alpha$ is Poisson structure.
   The homological vector field on $T^*[1]M$ is
  \beq
      Q=  2 \alpha^{\mu\nu} \eta_\nu \frac{\partial} {\partial x^\mu} + \partial_\mu \alpha^{\nu\rho} \eta_\nu \eta_\rho \frac{\partial}{\partial \eta_\mu}~,
  \eeq
    which gives rise to the Lie algebroid structure on $T^*M$ associated to a Poisson structure on $M$.

\begin{example}\label{example1} {\rm  ({\bf Lie algebroid})
 Consider  $L$ being a vector bundle over $M$ with the Lie algebroid structure described above. Then the dual bundle $L^*$ considered
  as a total manifold is equipped with a Poisson structure
  \beq\label{Poissonalgebd}
   \alpha (x, \lambda) = f_{AB}^C (x) \lambda_C ~\frac{\partial}{\partial \lambda_A} \wedge \frac{\partial}{\partial \lambda_B} +
  2  A^\mu_A (x) ~\frac{\partial}{\partial \lambda_A} \wedge \frac{\partial}{\partial x^\mu}~,
  \eeq
   where the $x$'s are coordinates on $M$ and the $\lambda$'s are coordinates on the fiber of $L^*$. Both $x$ and $\lambda$ are of degree zero.
    The corresponding symplectic manifold of degree 1 is $T^*[1]L^*$ with the symplectic
     structure of degree 1
     $$ \omega = d\eta_\mu dx^\mu + d j^A d\lambda_A~,$$
   where $\eta, j$ are the coordinates of degree 1.  From (\ref{Poissonalgebd}) it follows that the nilpotent Hamiltonian of degree 2 is given
    \beq
     \Theta =  f_{AB}^C (x) \lambda_C j^A j^B +  2 A^\mu_A (x) j^A \eta_\mu~.
    \eeq
 Thus Lie algebroid structure on $L$ can be encoded in terms of $(T^*[1]L^*, \omega, \Theta)$.}
\end{example}

Now let us discuss the graded symplectic manifolds of degree 2.
 The  symplectic (nonnegatively) graded symplectic manifold ${\cal M}$ of degree 2 corresponds
to vector bundle $E$ over $M$ with a fiberwise nondegenerate symmetric inner product $\langle~,~\rangle$
 (it can be of arbitrary signature).  For a given $E$, $\M$ is a symplectic submanifold of $T^*[2] E[1]$ corresponding
  to the isometric embedding $E \hookrightarrow E \oplus E^*$
    with respect to the canonical pairing on $E\oplus E^*$,  i.e. $e^a \rightarrow (e^a, g_{ab} e^b)$,  where $g_{ab}$
     is the constant fiber metric $\langle~,~\rangle$ written in a local basis of sections for $E$.
      Indeed ${\cal M}$ is a minimal symplectic realization of $E[1]$.
   In local Darboux coordinates $(x^\mu, p_\mu, e^a)$ of degree 0,2 and 1 respectively the symplectic structure is
\beq\label{symCOur747}
\omega = dp_\mu dx^\mu + \frac{1}{2} de^a  g_{ab} de^b~.
\eeq
 Any degree
 3 function would have the following general form
\bea\label{formcorudkkk}
\Theta=p_{\mu}A_a^{\mu}(x) e^a+\frac{1}{6}f_{abc} (x) e^ae^be^c~.
\eea
 As it has been shown in \cite{Roytenberg:2002nu} the solutions of the equation $\{ \Theta, \Theta\}=0$ correspond
  to Courant algebroid structures on $(E, \langle~, ~ \rangle)$
with the Courant--Dorfman bracket given by
$[\cdot,\cdot]=\{\{\cdot, \Theta\},\cdot\}$, where $\{~,~\}$ stands for the Poisson bracket on the symplectic manifold $\M$.
In  the expression (\ref{formcorudkkk})  the quantity $A_a^{\mu}$ and $-g^{da}f_{dbc}$ are interpreted as the anchor
and the structure function for Courant algebroid $E$, respectively.
We refer the reader to \cite{Roytenberg:2002nu} for detailed discussion of
degree 2 symplectic GrMflds and its relation to Courant algebroids.

\begin{example}\label{example2} {\rm  ({\bf  $TM\oplus T^*M$ Courant algebroid})
  The standard example of a Courant algebroid is the tangent plus cotangent bundle $TM \oplus T^*M$
   of a smooth manifold $M$. In this case the corresponding symplectic manifold of degree 2  is
${\cal M}=T^*[2]T^*[1]M$. The degree 0,1 subspace in this case is
$T^*[1]M\oplus T[1]M$. Pick local coordinates
$p_{\mu},v^{\mu},q_{\mu},x^{\mu}$ with degree 2,1,1 and 0 with the
metric induced from the natural pairing between $TM$ and $T^*M$.
The Hamiltonian function of degree 3 is
 $\Theta=P_\mu v^\mu$ which  induces a
vector field corresponding to the Hamiltonian lift of the de~Rham differential
\bea Q=\{\Theta,\cdot\}=v^{\mu}\frac{\partial}{\partial
x^{\mu}}+p_{\mu}\frac{\partial}{\partial q_{\mu}}~.\nn
\eea
If there is a closed 3-form $H$, then there exists another Hamiltonian function of degree 3
\beq\label{TstarThamilton}
\Theta = p_\mu v^\mu + \frac{1}{6} H_{\mu\nu\rho} v^\mu v^\nu v^\rho~,
\eeq
which gives rise to the twisted Courant structure on $TM\oplus T^*M$.}
\end{example}

\begin{example}\label{example3} {\rm  ({\bf Lie bialgebroid}) Consider a Lie algebroid $L$ and assume that the dual
 bundle $L^*$ is equipped with a Lie algebroid structure. The pair $(L, L^*)$ is called bialgebroid if
  $d_L$ is a derivation of the Schouten bracket on $\Gamma(\wedge^\bullet L^*)$.  For any bialgebroid $(L, L^*)$
   the vector bundle  $E=L\oplus L^*$ is naturally equipped with the structure of Courant algebroid \cite{weinstein}.  Thus we
    can apply the previous considerations. The graded manifold $T^*[2]L[1]$ is equipped with the symplectic structure
     of degree 2
\beq\label{syml3dkdk}
  \omega = dp_\mu dx^\mu + d\ell_A d\ell^A~,
  \eeq
  where $\ell^A$ are fiber coordinates\footnote{Through the paper we adapt the same notation for the fiber coordinates
 of a vector bundle and the sections of the dual bundle.} of degree 1 for  $L$ and $\ell_A$ are fiber coordinates of degree 1 for  $L^*$.
   The Hamiltonian of degree 3 has the same form as in (\ref{formcorudkkk}), but written in the basis adapted to
    the bialgebroid splitting $E= L\oplus L^*$. }
\end{example}

\begin{example}\label{example4} {\rm  ({\bf Lie algebroid}) Take a Lie algebroid $L$, then
 the vector bundle $E= L \oplus L^*$ can be regarded as bialgebroid with the
  trivial bracket and zero anchor on $L^*$. Thus $E= L \oplus L^*$ is
 equipped with Courant algebroid structure.
 The corresponding graded symplectic manifold is $T^*[2]L[1]$
  with the symplectic structure (\ref{syml3dkdk}).
  The corresponding Hamiltonian of degree 3 is
\bea &&\Theta=2p_{\mu}A_A^{\mu}(x) \ell^A
- f^A_{BC}(x) \ell_A \ell^B\ell^C~.\nn
\eea
The vector field $Q=\{\Theta,\cdot\}$ acts as the Lie algebroid differential  on
functions $f(x,\ell^A)$. In general it has an interpretation related
to the adjoint representation of $L$ \cite{abad-2009}.  }
\end{example}

\subsection{2D AKSZ model}

We may apply the AKSZ approach to the 2D case when the source manifold ${\cal N} = T[1]\Sigma_2$
with $\Sigma_2$ being a two-dimensional manifold. The target $\M = T^*[1]M$ is a symplectic manifold of
 degree 1 equipped with the Hamiltonian (\ref{Poissonham}). The space of fields defined as
 $$ {\rm Maps}(T[1]\Sigma_2,T^*[1]M)$$
 with odd symplectic structure
 \beq
   \omega_{BV} = \int\limits_{T[1]\Sigma_2} d^2\xi d^2\theta~ \delta \boldsymbol{\eta}_\mu \delta {\mathbf X}^\mu~.
 \eeq
  The corresponding BV action is written as
 \beq\label{PSM-AKSZ}
 S_{BV} = \int\limits_{T[1]\Sigma_2} d^2\xi d^2\theta \left ( \boldsymbol{\eta}_\mu D  {\mathbf X}^\mu +
  \alpha^{\mu\nu} ({\mathbf X}) \boldsymbol{\eta}_\mu \boldsymbol{\eta}_\nu \right )~,
 \eeq
  where we use  bold letters for the superfields corresponding to the coordinates on $T^*[1]M$, $\eta$ and $x$.
  This action is the BV formulation of the Poisson sigma model corresponding to the Poisson manifold $(M, \alpha)$.
   The action (\ref{PSM-AKSZ}) is a solution of
   a classical master equation if $\partial \Sigma_2 = \emptyset$, \cite{Cattaneo:2001ys}.  If $\partial \Sigma_2 \neq \emptyset$ then
    the following boundary conditions can be imposed
    $$ T[1] \partial \Sigma_2 ~\longrightarrow~ N^*[1] C~,$$
   where $C$ is a coisotropic submanifold of $M$. With these boundary conditions the requirements (\ref{BC1}) and (\ref{BC2})
    are satisfied \cite{Cattaneo:2003dp}.

In particular we are interested in the situation when the Poisson manifold is the dual bundle of a Lie algebroid $L$,
 see {\bf Example \ref{example1}}.  In this case the space of fields is
$$ T[1]\Sigma_2~\longrightarrow ~T^*[1] L^*$$
with the odd symplectic structure
\beq\label{2dsympls39999}
 \omega_{BV} =  \int\limits_{T[1]\Sigma_2} d^2\xi d^2\theta \left (  \delta \boldsymbol{\eta}_\mu \delta {\mathbf X}^\mu
  + \delta {\mathbf j}^A \delta \boldsymbol{\lambda}_A \right )~.
\eeq
The corresponding BV action is
\beq\label{2DLiealg}
 S_{BV} = \int\limits_{T[1]\Sigma_2} d^2\xi d^2\theta \left ( \boldsymbol{\eta}_\mu D  {\mathbf X}^\mu + {\mathbf j}^A D \boldsymbol{\lambda}_A
  +  f_{AB}^C ({\mathbf X}) \boldsymbol{\lambda}_C {\mathbf j}^A {\mathbf j}^B +  2 A^\mu_A ({\mathbf X}) {\mathbf j}^A \boldsymbol{\eta}_\mu \right)~,
\eeq
 where we use the obvious correspondence between superfields and the coordinates on $T^*[1]L^*$.
The action (\ref{2DLiealg}) satisfies the classical master equation if $\partial \Sigma_2 = \emptyset$. If $\partial \Sigma_2 \neq \emptyset$
 then the following boundary conditions can be imposed
 $$ T[1]\partial \Sigma_2 ~\longrightarrow~N^*[1] K^\perp~,$$
  where $K$ is a subalgebroid of $L$ and  $K^\perp\subset K^*$ is the annihilator of $K$:
\[
K^\perp_x = \{\alpha\in K^*_x :\alpha(v)=0\ \forall v\in K_x\}.
\]
   Let us remind that  a Lie subalgebroid $K$ of $L$ is a morphism of Lie algebroids $F: K \rightarrow L$, $f: C \rightarrow M$.
    such that $F$ and $f$ are injective immersions.
  It is easy to see that for boundary conditions labelled by subalgebroid of $L$ the conditions (\ref{BC1}) and
(\ref{BC2}) are satisfied, see \cite{Bonechi:2005sj} for similar analysis.

\subsection{3D AKSZ model}\label{subAKSZ3D}

 Having the general description of a graded symplectic manifold of degree 2 and the Hamiltonian
  function $\Theta$ of degree 3  we can write the AKSZ action for a Courant algebroid $E$. The space of fields
   is defined as
$$ T[1]\Sigma_3~\longrightarrow~\M ~,$$
where $\M$ is a symplectic submanifold of $T^*[2] E[1]$ which provides a minimal symplectic realization of $E[1]$.
The odd symplectic structure on the space of maps is
 \beq\label{sjdkk393939}
   \omega_{BV} =   \int\limits_{T[1]\Sigma_3} d^3\xi d^3\theta \left   ( \delta {\mathbf P}_{\mu}
   \delta {\mathbf X}^{\mu}+ \frac{1}{2} \delta {\mathbf e}^{a}g_{ab} \delta {\mathbf e}^b \right )~.
 \eeq
The BV action is
\bea\label{3Dcourantaksz}
S_{BV}=\int\limits_{T[1]\Sigma_3} d^3\xi d^3\theta \left   ( {\mathbf P}_{\mu}D{\mathbf X}^{\mu}+ \frac{1}{2} {\mathbf e}^{a}g_{ab} D {\mathbf e}^b
 - {\mathbf P}_{\mu}A_a^{\mu}({\mathbf X}) {\mathbf e}^a- \frac{1}{6}f_{abc} ({\mathbf X}) {\mathbf e}^a {\mathbf e}^b {\mathbf e}^c \right )
\eea
where we identify the superfields and the coordinates on $\M$ in an obvious way. The action (\ref{3Dcourantaksz}) satisfies the classical
 master equation if $\partial \Sigma_3 = \emptyset$.  If  $\partial \Sigma_3 \neq \emptyset$ then the additional boundary conditions
  should be imposed
  $$ T[1]\partial \Sigma_3~\longrightarrow~ {\cal L}~,$$
where ${\cal L}$ is a submanifold of $N^*[2] K[1]$ corresponding to the isometric embedding $E \hookrightarrow E \oplus E^*$
 (see previous discussion) and $K$ is a Dirac structure supported on a submanifold $C$ \cite{bursztyn}.
 A Dirac structure supported on a submanifold $i\colon C \hookrightarrow M$ is defined
  as a subbundle $K \subset i^* E = E|_C$ such that
  $K_x \subset E_x$ is maximally isotropic for all $x \in C$, $K$ is compatible with the anchor (i.e., $A(K) \subset TC$) and 
$[e_1, e_2]|_C \in \Gamma(K)$
   for any sections $e_1, e_2$ of $E$ such that $e_1|_C, e_2|_C \in \Gamma(K)$.

Let us illustrate the  general construction with a few  concrete examples.
 We start with the Courant algebroid structure over  $TM\oplus T^*M$.  The space of fields is described as
 follows
 $$ T[1]\Sigma_3 ~\longrightarrow~T^*[2]T^*[1] M$$
and the  BV action is
\bea\label{3DcourantakszstarTT}
S_{BV}=\int\limits_{T[1]\Sigma_3} d^3\xi d^3\theta \left   ( {\mathbf P}_{\mu}D{\mathbf X}^{\mu}+  \frac{1}{2}{\mathbf v}^\mu D {\mathbf q}_\mu
+ \frac{1}{2} {\mathbf v}^\mu D {\mathbf q}_\mu
 - {\mathbf P}_{\mu} {\mathbf v}^\mu- \frac{1}{6}H_{\mu\nu\rho} ({\mathbf X}) {\mathbf v}^\mu {\mathbf v}^\nu {\mathbf v}^\rho \right )~,
\eea
where we use the notations adapted to {\bf Example \ref{example2}}.  If $\partial \Sigma_3 \neq \emptyset$ then the possible
 boundary conditions would be
 $$  T[1]\partial \Sigma_3~\longrightarrow~ N^*[2] T[1]C~,$$
where $C$ is a submanifold and $H|_C=0$.  In this case $K= N^*C \oplus TC$ is an example of Dirac structure supported
 on $C$.  There is another way to construct a Dirac structure supported on $C$. Let us choose a two form on $C$, $B \in \Omega^2(C)$.
  Then applying the $B$-transform to $N^*C\oplus TC$ we obtain another bundle $e^B (N^*C\oplus TC)$ over $C$. It is easy to show that
   this gives rise to a Dirac structure with support over $C$ if $H|_C = dB$.  The pair $(C, B)$ with the condition $H|_C = dB$ has been discussed
    by Gualtieri \cite{Gualtieri:2003dx, Gualtieri:2007ng}, under the name of generalized submanifold.
      Using the local coordinates from {\bf Example \ref{example2}} adapted
     to submanifold $C$ we have the following description of the Lagrangian submanifold ${\cal L}$ in $T^*[2]T^*[1]M$
     $$ x^n=0~,~~~ v^n=0~,~~~q_i = B_{ij}(x) v^j~,~~~ p_i = -\frac{1}{2} \partial_i B_{jk} v^j v^k~,$$
where $n$ stands for the normal directions and $i, j, k$ for the tangential directions for $C$.  One can easily check that
 the Liouville form $\Xi= p_\mu dx^\mu + v^\mu dq_\mu$ and the Hamiltonian $\Theta$ (\ref{TstarThamilton}) vanish when restricted to ${\cal L}$
  provided that $H|_C = dB$.  Thus the appropriate boundary condition for the BV model would be
  $$   T[1]\partial \Sigma_3~\longrightarrow~ {\cal L}~,$$
   where ${\cal L}$ corresponds to a pair $(C, B)$ in a way described above.

   Now let us discuss the BV theory corresponding to {\bf Example \ref{example4}}. Again this is just a special case of Courant
    sigma model. Consider the space of fields described as
    $$ T[1]\Sigma_3 ~\longrightarrow~ T^*[2]L[1]~,$$
with the odd symplectic structure
\beq\label{syml3Dakkkkkk}
  \omega_{BV} = \int\limits_{T[1]\Sigma_3} d^3\xi d^3\theta \left   (\delta {\mathbf P}_\mu \delta {\mathbf X}^\mu +
   \delta \boldsymbol{\ell}_A \delta \boldsymbol{\ell}^A \right )
\eeq
and  the BV action given by the following expression
 \bea\label{63Dcoualgennskszstar}
S_{BV}=\int\limits_{T[1]\Sigma_3} d^3\xi d^3\theta \left   ( {\mathbf P}_{\mu}D{\mathbf X}^{\mu}+ \frac{1}{2}  \boldsymbol{\ell}_A D \boldsymbol{\ell}^A
+ \frac{1}{2}  \boldsymbol{\ell}^A D \boldsymbol{\ell}_A
 - 2 {\mathbf P}_{\mu} A_A^\mu ({\mathbf X}) \boldsymbol{\ell}^A + f^A_{BC}({\mathbf X})   \boldsymbol{\ell}_A \boldsymbol{\ell}^B
 \boldsymbol{\ell}^C  \right )
\eea
where our notations are adopted to {\bf Example \ref{example4}}.  If $\partial \Sigma_3 \neq \emptyset$ then the following
 boundary conditions should be imposed
 $$ T[1]\partial \Sigma_3 ~\longrightarrow~ N^*[2] K[1]~,$$
  where $K$ is a subalgebroid of $L$. It is straightforward to see that the conditions (\ref{BC1}) and (\ref{BC2}) are satisfied for
   this choice.

\section{AKSZ for Generalized Complex Manifolds}
\label{AKSZ_for_GCG}

In this section we discuss the description of generalized complex geometry (GCG) in terms of
graded symplectic manifolds. We apply this to the construction of the AKSZ action in 2D and 3D cases.

\subsection{Graded geometry for GCM}\label{subsectAKSZ2d}

Consider the Courant algebroid $E$ and associated to it the symplectic graded manifold ${\cal M}$
of degree 2. The Courant structure on $E$ is defined through the Hamiltonian function of degree 3
\beq
{\cal S}= p_{\mu}A_a^{\mu}(x) e^a+\frac{1}{6}f_{abc} (x) e^ae^be^c~,
\eeq
where from now on we use ${\cal S}$ to denote this concrete Hamiltonian.
 Consider the following function of degree 2 independent from $p$:
$$ J = \frac{1}{2} J_{ab}(x) e^a e^b~,$$
where by construction $J_{ab}= - J_{ba}$.  In \cite{grabowski} it has been observed
 that the function ${\cal S}$ and $J$ satisfy the relation
\beq\label{grabrelation}
 \{J,\{J, {\cal S}\}\}=-{\cal S}
\eeq
if and only if $J^a_{\ b} = g^{ac}J_{cb}$ defines the splitting of $E \otimes {\mathbb C} = L\oplus \bar{L}$ where
 $L$ is a maximally isotropic subbundle closed under the Courant bracket.
  One of the condition which follows from (\ref{grabrelation}) is $J^a_{\ b} J^b_{\ c} = -\delta^a_b$ and thus
  the subbundle $L$ is defined
  as $+i$ eigenbundle of $J^a_{\ b}$, thus $L^* = \bar{L}$.
    Although the interpretation of (\ref{grabrelation}) has been presented
   in \cite{grabowski} we prefer to give the details in the Appendix \ref{appendix1}.
     We find a number of useful formulas while
    investigating this relation.
   If we choose the Courant algebroid  $E=T^*M\oplus TM$,  then the splitting $(TM \oplus T^*M) \otimes {\mathbb C}=
 L \oplus \bar{L}$ with $L$ being a maximally isotropic involutive subbundle defines a generalized complex
structure (GCS) \cite{Gualtieri:2003dx, Gualtieri:2007ng}.  The Courant bracket restricted to $L$ becomes
a Lie bracket $[\cdot,\cdot]$ and thus $L$ is a complex Lie algebroid.

Next we observe that, if on ${\cal M}$ we have the functions ${\cal S}$ and $J$ with the the property (\ref{grabrelation}),
 then we can construct the function of degree 3
\beq\label{ssksww00slw}
 \Theta_{(\alpha, \beta)} = \alpha {\cal S} + \beta \{ J, {\cal S}\}~
\eeq
 which satisfies $\{\Theta, \Theta \}=0$ for arbitrary constants $\alpha$ and $\beta$.  If $\alpha$ and $\beta$ are real numbers,
  then there exists a symplectomorphism on ${\cal M}$ which connects $(\alpha^2 +\beta^2)^{1/2} {\cal S}$ with
   $\Theta_{(\alpha, \beta)}$. Namely $J$ gives rise to the flow
  $$ \partial_t \Theta(t) = \{ J, \Theta(t)\}~,$$
   which has the following explicit solution
   $$ \Theta (t) = (\alpha^2 + \beta^2)^{1/2} \left ( \cos t~ {\cal S} + \sin t ~\{J, {\cal S}\} \right )~.$$
   At $t=0$, $\Theta(t)$ corresponds to $(\alpha^2 +\beta^2)^{1/2} {\cal S}$. On the other hand if we choose $t$ to be such that
    $\cos t = \alpha (\alpha^2 +\beta^2)^{-1/2}$ and $\sin t = \beta (\alpha^2 +\beta^2)^{-1/2}$  then $\Theta(t)$ coincides
     with (\ref{ssksww00slw}). Therefore we do not get a new Hamiltonian of degree 3 if we deal with the real
      coefficients in (\ref{ssksww00slw}). Indeed any function  of degree 2 generates a symplectomorphism of ${\cal M}$
       \cite{Roytenberg:2002nu} and our particular function $J$  realizes the $U(1)$ action on ${\cal M}$.

To get something nontrivial we have to complexify our graded symplectic manifold ${\cal M}$ and allow
  complex coefficients in (\ref{ssksww00slw}). One can be easily convinced that, up to equivalence, the only non-trivial
  complex nilpotent Hamiltonians are
 \beq\label{newHamimpor}
  \Theta = {\cal S} + i \{J, {\cal S}\}
 \eeq
  and its complex conjugate. All other complex combinations of ${\cal S}$ and $\{J, {\cal S}\}$ do not give
   rise to anything new (it can be seen through the appropriate redefinitions).
  It is natural to choose the coordinates adapted to the splitting $E=L \oplus \bar{L}$.  The manifold
   ${\cal M}$ with the symplectic structure (\ref{symCOur747}) is symplectomorphic to $T^*[2] \bar{L}[1]$
    with the symplectic structure
    $$ \omega = d\tilde{p}_\mu dx^\mu + d\bar{\ell}_A d \ell^A~,$$
     where $\bar{\ell}_A$ are odd coordinate along a fiber of $\bar{L}$ and  $\ell^A$ along a fiber of $L$.
     Moreover the Hamiltonian (\ref{newHamimpor}) written in new coordinates becomes
  \beq\label{3399jjjs33sbb}
  \Theta=2\tilde{p}_{\mu}A_A^{\mu}(x) \ell^A
- f^A_{BC}(x) \bar{\ell}_A \ell^B\ell^C~.
\eeq
The proof of these statements  and further technical details are
  presented in Appendix \ref{appendix2}.  In the new coordinates the function
 $J$ looks particular simple $J = i \bar{\ell}_A \ell^A$.  The manifold $T^*[2] \bar{L}[1]$
has two natural gradings described by
\bea
&&\epsilon_1 = \tilde{p}_\mu \frac{\partial }{\partial \tilde{p}_\mu} + \bar{\ell}_A \frac{\partial }{\partial \bar{\ell}_A}~,\\
&&\epsilon_2 = \tilde{p}_\mu \frac{\partial }{\partial \tilde{p}_\mu} + \ell^A \frac{\partial }{\partial \ell^A}~,
\eea
where $\epsilon_1 + \epsilon_2$ corresponds to the original grading and
$\epsilon_1 - \epsilon_2$  is generated by $i J$. Thus the homological vector field for
 the Hamiltonian (\ref{newHamimpor}) comes from the splitting  of $\{{\cal S}, \cdot \}$ according to
  the grading defined by $\epsilon_1 - \epsilon_2$.

 Here we have discussed the complex case. However if in the relation (\ref{grabrelation}) the sign minus on
  the right hand side is replaced by plus then this corresponds to a real bialgebroid $E= L \oplus L^*$.
   A similar discussion with a few minor changes can be repeated for this case.

\subsection{AKSZ for 3D $\sigma$-model on GCM}

Using the discussion from the previous subsection, it is straightforward to construct the appropriate BV
 master action. Starting from the manifold ${\cal M}$ we define the space of maps as
 $$T[1]\Sigma_3~\longrightarrow~{\cal M}~.$$
Using the complex Hamiltonian (\ref{newHamimpor}) on ${\cal M}$, we construct the master action
 \bea\label{fullacion22}
S_{BV}&=& \int\limits_{T[1]\Sigma_3} d^3\xi d^3\theta \left   ( {\mathbf P}_{\mu}D{\mathbf X}^{\mu}+ \frac{1}{2} {\mathbf e}^{a}g_{ab} D {\mathbf e}^b -(\delta^a_b - i J^a_{\ b}({\mathbf X}))
{\mathbf e}^b A_a^{\mu} ({\mathbf X}) {\mathbf P}_{\mu}   \right .\nn\\
&&\left .-\frac{1}{6}f_{abc}(\mathbf{X}) {\mathbf e}^a {\mathbf e}^b
{\mathbf e}^c +\frac{i}{2}J^d_{\ a} ({\mathbf X}) f_{dbc} ({\mathbf X})  {\mathbf e}^a {\mathbf e}^b {\mathbf e}^c +
 \frac{i}{2}A^{\mu}_c({\mathbf X}) \partial_{\mu}J_{ab}({\mathbf X}) {\mathbf e}^a {\mathbf e}^b {\mathbf e}^c \right ) ~,
\eea
where we use the notations adapted to our discussion of geometry of ${\cal M}$.  The corresponding
 odd symplectic structure is (\ref{sjdkk393939}).
  The action (\ref{fullacion22}) satisfies the classical master equation if $\partial \Sigma _3 = \emptyset$.
   If $\partial \Sigma _3 \neq \emptyset$ then we have to impose the additional boundary conditions on
    the fields.  Recall from subsection \ref{subAKSZ3D} that the boundary
     conditions for  the Courant sigma model
       are specified by the Dirac structure $K$ supported on a submanifold $C$; $K$ gives rise
      to a Lagrangian submanifold ${\cal L}$ of ${\cal M}$ and ${\cal S}|_{\cal L}=0$. Now we have to
       see when $\{ J, {\cal S}\}|_{\cal L}=0$. The simplest way to get it is to  require that $J|_{\cal L}=0$.
        This follows from a simple property of symplectic geometry: if two functions vanish on a given Lagrangian
         then their bracket also vanishes on this Lagrangian\footnote{The simplest way to prove it is to perform a
          calculation of a bracket in the coordinates adapted to a Lagrangian submanifold.}.  The simplest way
           to achieve this is to require that $K_x \subset E_x$  is preserved under action of $J^a_{\ b}$ for all $x\in C$.
            Since $K_x$ is maximally isotropic it would imply that $J|_{\cal L}=0$. To summarize,
             boundary conditions for the action (\ref{fullacion22}) are labelled by  Dirac structures $K$
               supported on $C$ which are invariant under of the action of $J^a_{\ b}$. 

    As an illustration let us  consider $E= TM \oplus T^*M$.  In this case a solution of equation
     (\ref{grabrelation}) gives rise a the generalized complex structure. The action (\ref{fullacion22}) can be easily rewritten
     for this case.   As we discussed in subsection \ref{subAKSZ3D} for a submanifold $C$ and two-form $B \in \Omega^2(C)$, there   exists a Dirac structure supported over $C$  which we denoted  $K=e^B (N^*C\oplus TC)$.
As discussed above $e^B (N^*C\oplus TC)$ gives rise to the correct boundary condition if we require that
 it is invariant under the action of the generalized complex structure.
      This corresponds exactly to the definition of generalized complex submanifold
      suggested by Gualtieri \cite{Gualtieri:2003dx}. Thus  the boundary conditions for 3D AKSZ model
       are labelled by generalized complex submanifolds.

The manifold ${\cal M}$ is symplectomorphic to $T^*[2]\bar{L}[1]$. This induces  a symplectomorphism
 at the level of fields. Namely, the symplectic structure (\ref{sjdkk393939}) can be mapped to
 \beq
   \omega_{BV} =  \int\limits_{T[1]\Sigma_3} d^3\xi d^3\theta \left   (\delta \tilde{\mathbf P}_{\mu}
   \delta {\mathbf X}^{\mu}+  \delta \bar{\boldsymbol{\ell}}_A \delta \boldsymbol{\ell}^A \right )~,
 \eeq
  which is defined over the space of maps
  $$T[1]\Sigma_3~\longrightarrow~T^*[2]\bar{L}[1]~.$$
   The explicit formulas for the redefinitions of fields can be obtained from those  given in Appendix \ref{appendix2}.
Moreover by using the explicit manipulations in the Appendix,  action (\ref{fullacion22}) is
recast into the following
\bea
S_{BV}= \int\limits_{T[1]\Sigma_3} d^3\xi d^3\theta \left   (
\tilde{\mathbf P}_{\mu}D {\mathbf X}^{\mu}+\frac{1}{2}\bar{\boldsymbol{\ell}}_AD\boldsymbol{\ell}^A
+\frac{1}{2} {\boldsymbol{\ell}}^A D \bar{\boldsymbol{\ell}}_A
-2{\tilde {\mathbf P}}_{\mu}A_A^{\mu} ({\mathbf X}) \boldsymbol{\ell}^A
+f^A_{BC} ({\mathbf X}) \bar{\boldsymbol{\ell}}_A \boldsymbol{\ell}^B \boldsymbol{\ell}^C\right )
\label{BV_action}\eea
 where we kept all boundary terms arising in the redefinition.  This is the complex version of 3D AKSZ
  theory corresponding to the Lie algebroid.  In this formulation we analyze the boundary conditions
   as at the end of subsection \ref{subAKSZ3D}.  Moreover this discussion will be naturally compatible
    with the way we described boundary conditions as Dirac structure supported over $C$ which are invariant
     under $J^a_{\ b}$.

\subsection{AKSZ for 2D $\sigma$-model on GCM}

Since the generalized complex structure gives rise to a complex Lie algebroid $L$ we can
apply the construction from subsection \ref{subsectAKSZ2d}. The space of fields is defined as
$$T[1]\Sigma_2~\longrightarrow~T^*[1]\bar{L}~,$$
 where we regard $L$ as a formal complex Poisson manifold. The BV action is
 \beq
  S_{BV} = \int\limits_{T[1]\Sigma_2} d^2\xi d^2\theta \left ( \boldsymbol{\eta}_\mu D  {\mathbf X}^\mu + {\mathbf j}^A D \boldsymbol{\lambda}_A
  +  f_{AB}^C ({\mathbf X}) \boldsymbol{\lambda}_C {\mathbf j}^A {\mathbf j}^B +  2 A^\mu_A ({\mathbf X}) {\mathbf j}^A \boldsymbol{\eta}_\mu \right)~,
 \eeq
with the anchor and structure constants for $L$ (see Appendices for the explicit expressions). Obviously
 this model can be written for the case of $TM \oplus T^*M$.
 The boundary conditions in this model corresponds to Lie subalgebroids of L.  If we consider the case of generalized
  complex structure then the generalized complex submanifold $C$ gives rise to a Lie algebroid over $C$, which
   can be interpreted as a Lie subalgebroid of $L$; see \cite{Gualtieri:2007ng} for the details.
    Thus, generalized complex submanifolds
    give rise to the correct boundary conditions for this model.

\section{Reduction from 3D to 2D}
\label{reduction2D}

In this Section we discuss the relation between the 3D and
2D models introduced above. The consistent reduction is done
through the following observation.

\subsection{Separation of UV and IR (Losev's trick)}

Losev \cite{losev} suggested a framework for  dealing with  effective
theories within the BV framework. This idea was further developed and used in
\cite{Mnev:2006ch, Costello:2007ei, Mnev:2008sa}. Here we apply the idea of
 effective theory to the dimensional reduction of 3D AKSZ theory down to 2D
  AKSZ theory. We believe that this is the correct conceptual framework for the discussion
   of the dimensional reduction within BV formalism.

The idea is essentially very simple; let us sketch it.
Assume  that the BV manifold is of a
product structure $V=V_{UV}\times V_{IR}$, and the odd Laplacian is also
decomposed $\Delta=\Delta_{UV}+\Delta_{IR}$. Let $S_{BV}$ be a
BV action satisfying the quantum master equation $\Delta e^{-S}=0$ on $V$.
We shall refer to $V_{UV}$ as UV degrees of freedom and
to  $V_{IR}$ as IR degrees of freedom.
 We can 'integrate out' the UV degrees of freedom
and get an 'effective action' just as one would do for a normal
quantum field theory. More concretely, pick a Lagrangian
submanifold
${\cal L}\hookrightarrow V_{UV}$ and define the effective action on $V_{IR}$ as
\beq
e^{-S_{eff}} = \int\limits_{\cal L} e^{-S_{BV}}~.
\eeq
One can check that $S_{eff}$ satisfies the quantum master equation on $V_{IR}$
\beq
\Delta_{IR} e^{-S_{eff}} =  \int\limits_{{\cal L}}  \Delta_{IR} e^{-S}=\int\limits_{{\cal L}} \Delta e^{-S} -\int\limits_{{\cal L}} \Delta_{UV}~e^{-S}=0~,
\eeq
where the first term vanishes due to the master equation and the second term due to the integration over 
  a Lagrangian submanifold of a $\Delta_{UV}$-exact term.
Furthermore, assume that two choices of Lagrangian submanifolds ${\cal L}\hookrightarrow V_{UV}$ and
${\cal L}'\hookrightarrow V_{UV}$
are related by a gauge fixing fermion $\Psi$ 
such that $\Delta \Psi =0$, then
\bea
\int\limits_{{\cal L}'}e^{-S}-\int\limits_{{\cal L}}e^{-S}=\int\limits_{{\cal L}}\{\Psi,e^{-S}\}=\int\limits_{{\cal L}}
-\Delta(\Psi
e^{-S})-\Delta(\Psi)e^{-S}+\Psi\Delta(e^{-S})=-\Delta_{IR}\int\limits_{{\cal L}}\Psi
e^{-S}.\nn\eea
Thus the change of the gauge fixing in UV-sector leads to change in $e^{-S_{eff}}$ up to $\Delta_{IR}$-exact term.
These manipulations are well-defined if the BV manifold is finite dimensional. For the infinite dimensional
 manifold this construction is formal. For further details of construction the reader may consult \cite{Mnev:2008sa}. 

Using this trick, we  start from some solution to the classical
master equation and integrate out certain degrees of freedom.
The remaining effective action will also satisfy the classical
master equation.

\subsection{3D AKSZ model on $\Sigma_3 = \Sigma_2\times I$}

Using the idea of partial gauge fixing and integration in the UV-sector we will show that the 3D theory on $\Sigma_2
\times I$  corresponding to the Lie algebroid defined by (\ref{syml3Dakkkkkk}) and (\ref{63Dcoualgennskszstar})
is equivalent to the
 2D theory on $\Sigma_2$ corresponding to the same Lie algebroid  defined
  by (\ref{2dsympls39999}) and (\ref{2DLiealg}). This works provided the
  following boundary conditions
  \beq
   T[1]\partial \Sigma_3~\longrightarrow~ L[1]
  \eeq
   are imposed on the 3D theory.

Let us present the details of the derivation.  We take the 3D source to be $T[1]\Sigma_3=T[1](\Sigma_2\times I)$,
and name the even and odd coordinates along the interval $I$ as $(\theta^t,t)$. Expand all
superfields according to $\Phi(t,\theta^t)=\Phi(t)+\theta^t\Phi_t(t)$ and perform explicitly the $\theta_t$-integration.
 The odd symplectic structure (\ref{syml3Dakkkkkk}) becomes
\beq\label{UVIRsympsp}
 \omega_{BV} = - \int d^2 \xi d^2 \theta dt \left (- \delta {\mathbf P}_\mu \delta {\mathbf X}_t + \delta \boldsymbol{\ell}_A
  \delta \boldsymbol{\ell}^A_t + \delta {\mathbf P}_{t\mu} \delta {\mathbf X}^\mu + \delta \boldsymbol{\ell}_{tA} \delta
  \boldsymbol{\ell}^A \right )
\eeq
 and the BV action (\ref{63Dcoualgennskszstar})  is
\bea
 S_{BV} &=&  \int d^2 \xi d^2 \theta dt \left ( {\mathbf P}_\mu \partial_t {\mathbf X}^\mu + {\mathbf P}_{t\mu} D {\mathbf X}^\mu - {\mathbf P}_\mu D {\mathbf X}_t^\mu  - \frac{1}{2} \boldsymbol{\ell}_A \partial_t \boldsymbol{\ell}^A +  \frac{1}{2} \boldsymbol{\ell}_{tA}
 D \boldsymbol{\ell}^A  \right . \nn \\
 &&
 - \frac{1}{2}\boldsymbol{\ell}^A \partial_t \boldsymbol{\ell}_A +  \frac{1}{2}\boldsymbol{\ell}^A_{t}
 D \boldsymbol{\ell}_A 
 +\frac{1}{2} \boldsymbol{\ell}_A D\boldsymbol{\ell}^A_t 
 + \frac{1}{2}\boldsymbol{\ell}^A D\boldsymbol{\ell}_{tA} 
- 2 {\mathbf P}_{t\mu} A_A^\mu ({\mathbf X}) \boldsymbol{\ell}^A  \\
&& - 2 {\mathbf P}_{\mu} (\partial_\nu A_A^\mu ({\mathbf X})) {\mathbf X}^\nu_t \boldsymbol{\ell}^A
 -2 {\mathbf P}_{\mu} A_A^\mu ({\mathbf X}) \boldsymbol{\ell}_t^A +
  \partial_\mu (f^A_{BC}({\mathbf X}) ){\mathbf X}_t^\mu \boldsymbol{\ell}_A \boldsymbol{\ell}^B \boldsymbol{\ell}^C \nn \\
&&  \left .  +f^A_{BC}({\mathbf X}) \boldsymbol{\ell}_{tA} \boldsymbol{\ell}^B \boldsymbol{\ell}^C
   + 2f^A_{BC}({\mathbf X})  \boldsymbol{\ell}_A \boldsymbol{\ell}^B \boldsymbol{\ell}_t^C \right )~, \nn
\eea
 where now $D$ stands for the de~Rham along $T[1]\Sigma_2$.
 Since the symplectic structure (\ref{UVIRsympsp})  decomposes in two separate pieces we can choose
  the UV sector to correspond to $({\mathbf X}_t^\mu, {\mathbf P}_\mu, \boldsymbol{\ell}_t^A, \boldsymbol{\ell}_A)$.
 Next we choose the Lagrangian $\mathcal{L}$ in the UV sector as follows: ${\mathbf X}_t^\mu =0$,  $\boldsymbol{\ell}_t^A=0$. We get
\bea
 S_{BV}|_\mathcal{L} &=&  \int d^2 \xi d^2 \theta dt \left ( {\mathbf P}_\mu \partial_t {\mathbf X}^\mu + {\mathbf P}_{t\mu} D {\mathbf X}^\mu   - \frac{1}{2} \boldsymbol{\ell}_A \partial_t \boldsymbol{\ell}^A +  \frac{1}{2} \boldsymbol{\ell}_{tA}
 D \boldsymbol{\ell}^A  \right . \nn \\
 &&  \left.
 - \frac{1}{2}\boldsymbol{\ell}^A \partial_t \boldsymbol{\ell}_A 
 + \frac{1}{2}\boldsymbol{\ell}^A D\boldsymbol{\ell}_{tA} 
- 2 {\mathbf P}_{t\mu} A_A^\mu ({\mathbf X}) \boldsymbol{\ell}^A  
 +f^A_{BC}({\mathbf X}) \boldsymbol{\ell}_{tA} \boldsymbol{\ell}^B \boldsymbol{\ell}^C
    \right )~, \nn
\eea
We are now left with the integration over the remaining fields in the UV-sector: ${\mathbf P}_\mu$ and $\boldsymbol{\ell}_A$.
   Thus,  we end up with the following IR action
  \beq\label{ssjwkwk002202we}
 S_{BV}^{IR} =  \int d^2 \xi d^2 \theta dt \left ({\mathbf P}_{t\mu} D {\mathbf X}^\mu   + \boldsymbol{\ell}_{tA}
 D \boldsymbol{\ell}^A  + 2 A_A^\mu ({\mathbf X}) \boldsymbol{\ell}^A  {\mathbf P}_{t\mu} 
    +f^A_{BC}({\mathbf X}) \boldsymbol{\ell}_{tA} \boldsymbol{\ell}^B \boldsymbol{\ell}^C \right )~, 
\eeq
where the integration over ${\mathbf P}$ implements the condition $\partial_t {\mathbf X}^\mu=0$ and the integration 
 over $\boldsymbol{\ell_A}$ implements the condition $\partial_t \boldsymbol{\ell}^A=0$. There is a subtlety in the present 
  integration over ${\mathbf P}_\mu$ and  $\boldsymbol{\ell}_A$. Namely, if 
   ${\mathbf P}_\mu$ and  $\boldsymbol{\ell}_A$ had zero modes ($t$-independent pieces) then we 
    would not be able to integrate them completely.  This is why we need  boundary conditions that 
     imply the absence of zero modes: namely,  
\bea 
P_\mu|_{T[1]\partial\Sigma_3}= \boldsymbol{\ell}_A|_{T[1]\partial\Sigma_3}=0~,\nn\eea
which are also the correct conditions from the points of view of the AKSZ construction. 
  The fields ${\mathbf X}^\mu$ and $\boldsymbol{\ell}^A$ are $t$-independent and the other fields
   ${\mathbf P}_{t\mu}$ and $\boldsymbol{\ell}_{tA}$ enter the action (\ref{ssjwkwk002202we}) linearly. 
    Therefore upon the following identification 
    \beq
    {\mathbf X}^\mu ={\mathbf X}^\mu~,~~~~\boldsymbol{\eta}_\mu = \int\limits_{I} dt~ {\mathbf P}_{t\mu}~,~~~
    {\mathbf j}^A = \boldsymbol{\ell}^A~,~~~\boldsymbol{\lambda}_A = \int\limits_{I} dt ~\boldsymbol{\ell}_{tA}~,
    \eeq
 the 3D action (\ref{ssjwkwk002202we}) collapses to the 2D theory given by (\ref{2DLiealg}). 

We have shown that the 3D AKSZ theory for a Lie algebroid on $\Sigma_2 \times I$ can be reduced to the
2D AKSZ theory for the same Lie algebroid, provided that the specific boundary conditions are imposed.
 We would like to stress that for the case $\Sigma_2 \times S^1$ the reduction would not work 
  properly due to the presence of zero modes.  

Here we have discussed the reduction for the real model. The reduction for the complex Lie algebroid works
 in exactly the same way and all expressions remain true modulo notations. 

\section{Summary}
\label{summary}

A Lie algebroid $L$ can  be encoded by saying that $L[1]$ is equipped with a homological vector field 
of degree 1. We considered two possible Hamiltonian lifts of this vector field, for the symplectic 
 manifold $T^*[1] L^*$ of degree 1 and for the symplectic manifold $T^*[2]L[1]$ of degree 2.  
  Using the AKSZ construction, these two lifts give rise to 2-dimensional and 3-dimensional topological 
   field theories respectively. We also discussed the allowed boundary conditions for these theories.
   Moreover, we have shown that the 3-dimensional theory on $\Sigma_2 \times I$ 
    reduces to the 2D theory on $\Sigma_2$,  upon specific boundary conditions. 

 A generalized complex structure is a complex Lie algebroid $L$ with the additional property that $\bar{L}= L^*$. 
  Thus, all our formal considerations are equally applicable to the case of a generalized complex structure. 
   One can show that our 2D theory with  generalized complex structure corresponding to an ordinary complex
    structure is, upon  gauge fixing,  equivalent to the B-model \cite{Witten:1991zz},  while the 2D theory for a symplectic structure is equivalent
     to the A-model \cite{Witten:1991zz}.  The more general 2D models on a generalized K\"ahler manifold 
      should correspond to a topological twist 
      of the $N=(2,2)$ nonlinear sigma model
     \cite{Kapustin:2004gv, Hull:2008de}.  We will present the detailed analysis of the gauge 
      fixing for these models  in a forthcoming work \cite{CMQZ}. 

\bigskip\bigskip

\noindent{\bf\Large Acknowledgement}:
\bigskip

\noindent
We are deeply grateful to Francesco Bonechi and Pavel Mn\"ev for the numerous useful discussions. 
The research of M.Z. was
supported  by VR-grant 621-2008-4273.
A.S.C. has been
partially supported by SNF Grant
200020-121640/1, by
the European Union through the FP6 Marie Curie RTN ENIGMA (contract
number MRTN-CT-2004-5652), and by the European Science Foundation
through the MISGAM program.

\appendix

\section{Integrability of generalized complex structure}\label{appendix1}

Grabowski \cite{grabowski}  suggested a description of  generalized complex geometry in terms of graded manifolds. 
Here we review this and provide a number of useful relations. 

 Following Roytenberg \cite{Roytenberg:2002nu} we describe a Courant algebroid $E$ in terms of a graded
  symplectic manifold ${\cal M}$
  of degree 2 with  Hamiltonian ${\cal S}$ of degree 3.  The manifold ${\cal M}$ is the minimal symplectic realization of 
   $E[1]$    and  in local Darboux coordinates $(x^\mu, p_\mu, e^a)$ of degree 0,2 and 1 respectively 
   the symplectic structure is
\beq\label{symCOur747lllll}
\omega = dp_\mu dx^\mu + \frac{1}{2} de^a  g_{ab} de^b~,
\eeq
where $g_{ab}$ is the constant fiber metric $\langle~,~\rangle$ written in a local basis of sections for $E$:
 $e^a$ are the fiber odd coordinates on $E$ which transform as sections of $E^*$. We use the metric $g_{ab}$
to raise and lower the indexes, thus relating  $E$ and $E^*$. 
Any degree
 3 function will have the following general form
\bea\label{formcorudkkklll}
\Theta=p_{\mu}A_a^{\mu}(x) e^a+\frac{1}{6}f_{abc} (x) e^ae^be^c~.
\eea
 As it has been shown in \cite{Roytenberg:2002nu}, the solutions of the equation $\{ \Theta, \Theta\}=0$ correspond
  to Courant algebroid structure on $(E, \langle~, ~ \rangle)$
with the Courant--Dorfman bracket given by
$[\cdot,\cdot]=\{\{\cdot, \Theta\},\cdot\}$, where $\{~,~\}$ stands for  the Poisson bracket on the symplectic manifold $\M$.
In  the expression (\ref{formcorudkkklll})  the quantity $A_a^{\mu}$ and $-g^{da}f_{dbc}$ are interpreted as the anchor
and the structure functions for the Courant algebroid $E$, respectively. Equivalently we can discuss the Courant algebroid 
structure on $E^*$.  The Courant--Dorfman brackets for coordinates  are given by  the 'structure functions'
  $f^{ab}{}_c$ to be $[e^a,e^b]=\{\{e^a,{\cal S}\},e^b\} = - f^{ab}{}_c e^c$ or equivalently 
   $\langle [e^a,e^b],e^c\rangle=\{\{\{e^a, {\cal S}\},e^b\},e^c\}=- f^{abc}$.
   Next   we define a function of degree 2 which is independent of $p$ as follows
   $J = \frac{1}{2} J_{ab}(x) e^a e^b$, 
    such that $J^a_{\ b}(x) = g^{ac} J_{cb} (x): E \rightarrow E$ is interpreted as an endomorphism of $E$ and 
     $J_a^{\ b}(x) = J_{ac} (x) g^{cb} : E^* \rightarrow E^*$ as an endomorphism of $E^*$. 
We want to study the relation    $\{J,\{J, {\cal S}\}\}=-{\cal S}$.  Expand out the brackets
\bea
\{J,{\cal S}\}&=&\{\frac{1}{2}J_{ab}(x) e^ae^b, p_{\mu}A^{\mu}_a(x)e^a + \frac{1}{6} f_{abc} e^a e^b e^c \}=\nn \\
&& -\frac{1}{2}(\partial_{\mu}J_{ab}(x)) e^a e^b e^cA_c^{\mu}(x) +e^cJ_{ca}(x)\left[A_b^{\mu}(x) p_{\mu}
+\frac{1}{2}f_{bde}(x) e^de^e\right]g^{ab}~,\nn\\
\{J,\{J, {\cal S}\}\}
&=&- (\partial_{\mu}J_{ad}(x))e^a e^d e^cJ_c^{\ b}(x)A_b^{\mu}(x) -(e^fJ_f^{\
h}(x) )(\partial_{\mu}J_{hk} (x))e^k e^aA_a^{\mu}(x)\nn\\%
&&+e^fJ_f^{\ h}(x) J_h^{\ b} (x) \bigg[A_b^{\mu}(x) p_{\mu}+\left(\frac{1}{6}+\frac{1}{3}\right)f_{bde}(x)e^de^e\bigg]\nn \\
&&-e^fJ_f^{\
h}(x)e^cJ_c^{\ b}(x)f_{bhe}(x)e^e~.\nn\eea%
Thus the condition $\{J,\{J, {\cal S}\}\}=-{\cal S}$ requires  
\bea
&&J_a^{\ b} (x) J_b^{\ c} (x) = - \delta_a^c ~, \label{djd99393939} \\
&& N(J)_{abc}\equiv -(\partial_{\mu}J_{ab}(x))J_c^{\
d} (x)A_d^{\mu}(x) -J_a^{\ d}(x) (\partial_{\mu}J_{db}(x))A^{\mu}_c (x) \nn \\
&&-\frac{1}{3}f_{abc}(x) -J_a^{\
e}(x)J_b^{\ d}(x)f_{dec}(x) +\textrm{cyclic in}(abc)=0 ~,\label{Intgrab}\eea
where by construction $N_{abc} \in \wedge^3 E$.
Condition (\ref{djd99393939}) implies that $E$ can decomposed into the sum of two maximally 
 isotropic spaces, namely $E\otimes \mathbb{C} = L \oplus \bar{L}$. We introduce the projection operators
$$\Pi_{\pm b}^a (x) = \frac{1}{2} \left ( \delta^a_b \pm i J^a_{\ b} (x) \right)~, $$
such that $\Pi_-$ projects $L$ and $\Pi_+$ its complex conjugate, $\bar{L}$.  By construction $L$ (and $\bar{L}$) are 
 maximally isotropic with respect to $g$, since $J_{ab}(x) = - J_{ba}(x)$. 
Next we show that (\ref{Intgrab}) gives the integrability
condition which states that $L$ is involutive under the Courant-Dorfman bracket $[\cdot,\cdot]$, that is
$$ \Pi_+ [ \Pi_- e^a, \Pi_- e^b]=0~.$$
The real part of this expression gives
\bea 
[e^a,e^b]-[J^a_{\ c}(x) e^c,J^b_{\ d}(x)e^d]+J[e^a,J^b_{\
d}(x)e^d]+J[J^a_{\ c}(x) e^c,e^b]=0~,\label{Itgb_temp1}\eea 
where by $J$ we understand the endomorphism of $E$. 
Using the relations 
\bea
[e^a,J^b_{\ d}(x) e^d]&=&\{\{e^a,{\cal S}\},J^b_{\ d}(x) e^d\}\nn\\
&=&A^{a\mu}(x)(\partial_\mu J^b_{\ d}(x))  e^d+J^b_{\ d}(x)[e^a,e^d]~,\nn \\
\left[J^a_{\ c}(x) e^c,e^b\right]&=&\{\{J^a_{\ c}(x) e^c, {\cal S}\},e^b\}\nn\\
&=&-A^{b\mu} (x) (\partial_\mu J^a_{\ c} (x))  e^c-J^a_{\
c}(x) [e^b,e^c]+e^dA_d^{\mu}(x) (\partial_{\mu}J^{ab}(x)) ~,\nn\eea
and the 'structure constant' $f^{ab}{}_c$ 
the expression (\ref{Itgb_temp1}) can be rewritten as
\bea 
&&- f^{ab}{}_ce^c+J^d_{\ c}(A^{a\mu}\partial_\mu J^b_{\
d})e^c -J^b_{\ d}f^{ad}{}_eJ^e_{\ c}e^c-J^d_{\
c}e^cA^{b\mu}(\partial_\mu J^a_{\ d}) +J^a_{\ c}f^{bc}{}_dJ^d_{\
e}e^e+J^d_{\ c}A_d^\mu \partial_\mu J^{ab}e^c\nn
\eea%
This can be further massaged into%
\bea A^{a{\mu}}(\partial_{\mu}J^b_{\
d})J^{dc}+J^{dc}A_d^{\mu}(\partial_{\mu}J^{ab})+\textrm{(cyclic in
a,b,c)} -f^{abc} -J^{[b}_{\ \ d}f^{a]d}{}_eJ^{ec}-J^b_{\ d}J^a_{\
e}f^{dec}=0\nn\eea%
and thus the integrability condition becomes%
\bea A_a^{\mu}(\partial_{\mu}J_{bd})J^d_{\ c}+J^d_{\
c}A_d^{\mu}(\partial_{\mu}J_{ab})+\left[-\frac{1}{3}f_{abc}+J^d_{\
b}J^e_{\ c}f_{ade}\right]+\textrm{(cyclic in a,b,c)}=0~,\nn\eea%
which coincides with (\ref{Intgrab}). Since the
sections of $L$ are now involutive under the Courant bracket
(which is antisymmetric when restricted to $L$), $L$ and likewise
$\bar L$ defines a \emph{Lie algebroid} structure. 

\begin{example}\label{example5} {\rm  
Consider the standard Courant algebroid structure on $TM\oplus T^*M$ and the corresponding graded 
 symplectic manifold $T^*[2]T^*[1]M$ from {\bf Example \ref{example2}}. 
Consider the function of degree 2 of the form
$$J=J^{\mu}_{\ \nu}(x)q_{\mu}v^{\nu}~.$$
 Condition (\ref{djd99393939}) says that $J^\mu_{\ \nu}$ is almost complex structure
  and the condition (\ref{Intgrab}) becomes
\bea J_{\rho}^{\ \sigma}\partial_{\sigma}J_{\mu}^{\
{\nu}}+J_{\mu}^{\ {\sigma}}\partial_{\sigma}J^{\nu}_{\
{\rho}}+J_{\mu}^{\ {\sigma}}\partial_{\rho}J_{\sigma}^{\
{\nu}}+J^{\nu}_{\ {\sigma}}\partial_{\mu}J^{\sigma}_{\
{\rho}}=0~,\eea%
which we recognize as the standard Nijenhuis tensor: $J^{\sigma}_{\
[\rho}\partial_{\sigma}J^{\nu}_{\ \mu]}-J^{\nu}_{\
{\sigma}}\partial_{[\rho}J^{\sigma}_{\ \mu]}=0$. Thus we end up with the standard complex structure on $M$.}
\end{example}

\begin{example} {\rm 
Take another example of function of degree 2 on $T^*[2]T^*[1]M$
$$J=\frac{1}{2}(\omega_{\mu\nu}(x){v}^{\mu}{v}^{\nu}+\omega^{\mu\nu}(x){q}_{\mu}{q}_{\nu})~.$$
 Condition (\ref{djd99393939}) implies that $\omega_{\mu\nu} \omega^{\nu\lambda} =\delta_\mu^\lambda$ and 
  condition  (\ref{Intgrab}) becomes
$\omega^{\sigma[\mu}\partial_{\sigma}\omega^{{\nu\rho}]}=0$.
Thus, $\omega_{\mu\nu}$ is a closed non-degenerate two-form, symplectic structure.}
\end{example}

\begin{example}{\rm 
On $T^*[2]T^*[1]M$ the general form of function of degree 2 independent from $p$ is 
$$ J = \frac{1}{2} L_{\mu\nu} (x) v^\mu v^\nu + J^\mu_{\ \nu}(x) q_\mu v^\nu + \frac{1}{2}  P^{\mu\nu}(x) q_\mu q_\nu~.$$
 By plugging this into (\ref{djd99393939}) and (\ref{Intgrab}) we get the conditions for a generalized complex 
  structure on $TM \oplus T^*M$ which were analyzed in \cite{Lindstrom:2004iw, Crainic:2004ic}. We would like 
   to stress that the language of graded symplectic manifolds allows one to obtain those complicated conditions by performing
    rather simple calculations.}
\end{example}

We have another useful observation regarding the integrability of $J^a_{\ b}$.
Define 
$\partial_c^+$ by
$$\partial_c^+V^a\equiv
A_c^{\mu}\partial_{\mu}V^a+ \frac{1}{3}f^{a}_{\ bc}V^b~.$$
Then
$N_{abc}$ is in fact the (3,0)+(0,3) part of $\partial_{[c}^+J_{ab]} \in \wedge^3 E$.
Let us check this explicitly 
\bea &&(\partial^+_{c'}J_{a'b'}) \Pi^{a'}_{-a} \Pi^{b'}_{-b} \Pi^{c'}_{-c}+\textrm{(cyc in
abc)}=-(\partial^+_{c'}J^{a'}_{\ a}) \Pi_{-a'b} \Pi^{c'}_{-c}+\textrm{(cyc in abc)}\nn\\%
&=&-\frac{1}{4}\left[\partial^+_cJ_{ba}-J^{c'}_{\
c}(\partial^+_{c'}J^{a'}_{\
a})J_{a'b}\right]+\frac{i}{4}\left[J_{a'b}\partial^+_cJ^{a'}_{\
a}+J^{c'}_{\ c}\partial^+_{c'}J_{ba}\right]+\textrm{(cyc in abc)}\nn\\%
&=&-\frac{1}{4}\left[\partial^+_cJ_{ba}+J^{\ d}_cJ_a^{\
f}(\partial^+_dJ_{fb})\right]+\frac{i}{4}\left[J_a^{\
d}\partial^+_cJ_{db}+J_c^{\
d}\partial^+_dJ_{ab}\right]+\textrm{(cyc in abc)}\nn\\%
&=&\frac{1}{4}J_c^{\ d}N_{abd}-\frac{i}{4}N_{abc}~.\nn\eea%
Thus, the integrability condition says that $\partial_{[c}^+J_{ab]}$ is of type
(2,1)+(1,2). This reinterpretation of integrability is analogous to the description of the integrability 
of the almost complex structure $J$ on the Hermitian manifold $(J, g)$. The almost complex structure is integrable 
 if and only if $d\omega$ is of a type (2,1)+(1,2), where $\omega = gJ$.  

\section{Change of coordinates}\label{appendix2}

Consider the symplectic graded manifold ${\cal M}$ of degree 2 associated to a vector 
bundle $E$ with  fiberwise nondegenerate symmetric inner product $\langle~,~\rangle$. 
 In local Darboux coordinates the symplectic structure is given by (\ref{symCOur747lllll}). 
 Now assume that we have endomorphisms $J^a_{\ b}$ of $E$ such that 
  $J^a_{\ b} J^b_{\ c} = - \delta^a_c$ and $J_{ab} = g_{ac} J^c_{\ b} = - J_{ba}$. 
   Thus, this endomorphism defines a splitting of $E$ into two maximally isotropic 
    subbundles, $E \otimes {\mathbb C} = L \oplus \bar{L}$.   Then the symplectic manifold ${\cal M}$
     is symplectomorphic to $T^*[2] L[1] = T^*[2] \bar{L}[1]$.  Let us show this explicitly. 

Introduce the vielbein $F_A^a$ which can simply be understood as the $i
$-eigenvector of the endomorphism $J^a_{\ b}$ labelled by index $A$. We
lower and raise the Euclidean indices with the pairing $g_{ab}$
and its inverse $g^{ab}$. The following properties of the vielbein
follow from their interpretation as the eigenvectors of $J^a_{\ b}$
\bea J^a_{\ b}F^b_A=iF^a_A;&&J^a_{\ b}F^{Ab}=-iF^{Aa}\ \ \textrm{definition}\nn\\%
F^A_aF_{Bb}g^{ab}=\delta^A_B;&&F^A_aF^B_bg^{ab}=F^a_AF_B^bg_{ab}=0\
\ \textrm{orthonormality}\nn\\%
F_A^aF^A_b=\Pi^a_{-b}&&F^{Aa}F_{Ab}=\Pi^a_{+b}\ \ \
\textrm{completeness}\nn\eea%
We introduce the following new coordinates
\bea &&\ell^A=F^A_ae^a;\ \bar\ell_A=F_{Aa}e^a\nn\\%
&&\tilde
p_{\mu}=p_{\mu}+\frac{1}{2}(\partial_{\mu}F_{Aa})F^a_B\ell^A\ell^B
+\frac{1}{2}(\partial_{\mu}F^A_a)F^{Ba}\bar\ell_A\bar\ell_B
-F^A_a(\partial_{\mu}F_B^a)\bar\ell_A\ell^B~,\nn\eea%
which are adopted to the splitting $E \otimes {\mathbb C} = L \oplus \bar{L}$.
The symplectic form  (\ref{symCOur747lllll}) goes to   $\omega=d
\tilde{p}_{\mu} d x^{\mu}+ d\bar\ell_A d\ell^A$. The
 Liouville form changes as follows
\bea 
\Xi=p_{\mu}dX^{\mu}+\frac{1}{2}e^a g_{ab} de^b\Rightarrow\tilde
p_{\mu}dx^{\mu}+\frac{1}{2}\bar\ell_Ad \ell^A+\frac{1}{2} \ell^A d\bar\ell_A~.\nn\eea%

 Next assume that the endomorphism $J^a_{\ b}$ is integrable in sense we have discussed in 
  the previous Appendix. 
We can give the integrability condition more concisely in this new
basis. Recall that
$N_{abc}=(\partial^+_{[a}J_{bc]})^{(3,0)+(0,3)}$, we have%
\bea
0&=&\frac{1}{2}F_A^aF_B^bF_C^c(\partial^+_{[a}J_{bc]})=F_A^aF_B^bF_C^c(\partial_aJ_{bc}-\frac{1}{3}f_{a\
b}^{\ d}J_{dc}+\frac{1}{3}f_{a\ c}^{\ d}J_{db})+\textrm{cyc in
}\small{ABC}\nn\\%
&=&2i((\partial_AF_C^c)F_{Bc}+\frac{1}{3} F_A^a F_B^b F_C^c f_{abc})+\textrm{cyc
in }\small{ABC}~, \label{}  \eea%
where we used the following notations:  $\partial_A:=F_A^aA_a^{\mu}\partial_{\mu}$, 
$\partial^A:=F^A_aA^{a\mu}\partial_{\mu}$. Using this identity we can show that 
\bea [\ell^A,\ell^B]&=&\{\{ p_\mu A_a^\mu e^a + \frac{1}{6} f_{abc} e^a e^b e^c,F^A_ae^a\},F_b^Be^b\}\nn\\%
&=&\ell^C\big(F_C^a(\partial_{[a}F_{c]}^A)F^{Bc}+F_C^b\partial^AF_b^B+ F_C^c F^{Bb} F^{Aa} f_{cba} \big )\nn\eea%
 and
similarly the other combination%
\bea[\bar\ell_A,\bar\ell_B]
=\bar\ell_C\big(F^{Ca}(\partial_{[a}F_{c]A})F_B^c+F^C_c\partial_AF^c_B+ F^{Cc}  F^{b}_B F^{a}_A f_{cba}  \big)~.\nn\eea%
In this new basis the corresponding Hamiltonian of degree 3  takes very simple form%
\bea 
{\cal S} + i \{J, {\cal S}\} &=& 2({\Pi}_-e)^a A^\mu_a p_\mu+\frac{1}{3}({\Pi}_-e)^a f_{abc}e^be^c-\frac{i}{2}(\partial^+_c
J_{ab})e^ae^be^c\nn\eea%
By using the fact $\partial^+_{[c} J_{ab]}$ is (2,1) and (1,2), we can
simplify the last term in this expression.
Thus, we can finally rewrite the Hamiltonian in the following form 
\bea
{\cal S} + i \{ J, {\cal S}\} =
2\ell^CA_C^{\mu}\tilde{p}_{\mu}-\ell^A\ell^B\bar\ell_Cf^C_{AB}~,\nn\eea
 where we use the data for the Lie algebroid $L$
 $$ A_C^\mu = F_C^a A_a^\mu~,$$
$$ f_{AB}^C = F^{Ca}(\partial_{[a}F_{c]A})F_B^c+F^C_c\partial_AF^c_B+ F^{Cc}  F^{b}_B F^{a}_A f_{cba} ~.$$

Here we performed the calculations for the complex bialgebroid $(L, \bar{L})$. The generalization for 
the real bialgebroid is a straightforward modifications of the present calculation.

\end{document}